\documentclass{article}
\usepackage[final]{graphics}

\newcommand{\slL}{\raise.15ex\hbox{$/$}\kern-.53em\hbox{$L$}}
\newcommand{\slP}{\raise.15ex\hbox{$/$}\kern-.53em\hbox{$P$}}
\newcommand{\slR}{\raise.15ex\hbox{$/$}\kern-.53em\hbox{$R$}}
\newcommand{\slQ}{\raise.15ex\hbox{$/$}\kern-.53em\hbox{$Q$}}
\newcommand{\slK}{\raise.15ex\hbox{$/$}\kern-.53em\hbox{$K$}}
\newcommand{\slD}{\raise.15ex\hbox{$/$}\kern-.7em\hbox{$D$}}
\newcommand{\slS}{\raise.15ex\hbox{$/$}\kern-.7em\hbox{$S$}}
\newcommand{\slA}{\raise.15ex\hbox{$/$}\kern-.7em\hbox{$A$}}
\newcommand{\slB}{\raise.15ex\hbox{$/$}\kern-.7em\hbox{$B$}}
\newcommand{\slC}{\raise.15ex\hbox{$/$}\kern-.7em\hbox{$C$}}
\newcommand{\slSigma}{\raise.15ex\hbox{$/$}\kern-.53em\hbox{$\Sigma$}}
\newcommand{\slcalP}{\raise.15ex\hbox{$/$}\kern-.63em\hbox{$\cal P$}}

\font\tenimbf=cmmib10 at 12pt
\font\sevenimbf=cmmib10 at 7pt
\font\fiveimbf=cmmib10 at 5pt
\newfam\imbf
\textfont\imbf=\tenimbf
\scriptfont\imbf=\sevenimbf
\scriptscriptfont\imbf=\fiveimbf
\def\imb{\fam\imbf\tenimbf}

\begin{document}
\date{February 25, 2000}
\begin{titlepage}
\title{\bf{Anomalous processes\\ at high temperature and density\\ 
in a 2-dimensional linear $\sigma$ model}}
\author{Fran\c cois~Gelis$^{(1)}$, Michel H.G.~Tytgat$^{(2)}$}
\maketitle

\begin{center}
\begin{enumerate}
\item Brookhaven National Laboratory,\\
Physics Department, Nuclear Theory,\\
Upton, NY 11973-5000, USA
\item CERN, Theory Division\\
CH-1211 Geneva 23, Switzerland
\end{enumerate}
\end{center}

\begin{abstract} We use the 2-dimensional $\sigma$  model as a toy 
  model to study the behavior of anomalous amplitudes in the limit
  where the constituent quark mass is small. Symmetry arguments tell
  that the $\pi^o\to\gamma$ amplitude should vanish if $m\to 0$, but
  we show that this conclusion is spoiled by infrared singularities.
  When a proper regularization (resummation of a thermal mass, for
  instance) is taken into account, this amplitude vanishes as
  expected. We also study the amplitude $\pi^o\sigma\to\gamma$ and
  show that it does not vanish in the same limit.
\end{abstract}
\vskip 4mm \centerline{\hfill CERN-TH/2000-066, BNL--NT--00/3, hep-ph/0002270} \vfill
\thispagestyle{empty}
\end{titlepage}

\section{Introduction}
In the past years, many works have been devoted to the study of
anomalous processes at finite temperature and density, and in
particular near the chiral symmetry restoration. In particular,
Pisarski \cite{Pisar8,Pisar9} concluded that the neutral pion decay
amplitude $\pi^o\to2\gamma$ vanish when chiral symmetry is
restored\footnote{This does not contradict the well established fact
  that the coefficient of the axial anomaly is temperature independent
  (see \cite{ItoyaM1,DolanJ1,BaierP1,NicolA2,Smilg3} for instance).
  Indeed, it has been shown in \cite{PisarT3,PisarT4,PisarTT1} that
  the relationship between the anomaly and the amplitude can be
  modified by the existence of an additional 4-vector $U_\mu$ (the
  4-velocity of the plasma in the observer's frame) that can enter the
  general form of thermal amplitudes.}.  This conclusion is based on a
direct calculation of the corresponding diagrams at finite temperature
(and zero density) in the imaginary time formalism, and on symmetry
considerations that forbid certain couplings in the symmetric phase.
An additional conclusion was that this decay might be replaced by
$\pi^o\sigma\to 2\gamma$ in the chirally symmetric phase (which is
allowed by the same symmetry argument), as indicated by a calculation
of the box diagram at finite temperature.

This result has been confirmed by Baier, Dirks and Kober
\cite{BaierDK1} and by Salcedo \cite{Salce1} using functional
approaches in which one integrates out the fermions at the level of
the the generating functional. Salcedo in \cite{Salce1} gives also
general arguments according to which $\pi^o\to2\gamma$ (or
$\pi^o\to\gamma$ in two dimensions) should vanish in the chiral phase,
and be replaced by amplitudes involving the $\sigma$ meson. From a
technical perspective, the common point of all these studies is the
use at some stage of the imaginary time formalism, in the limit of
vanishing external momenta.

In another study, one of us \cite{Gelis6} studied how the neutral pion
decay amplitude depends on the kinematical configuration of the
external legs in order to explain the discrepancies found between
\cite{Pisar8,Pisar9} and calculations performed in the real time
formalism by \cite{ContrL1,NicolA1,GuptaN1}. To that purpose, the
$\pi^o\gamma\gamma$ amplitude has been calculated at finite $T$ in the
real time formalism, in the limit of small external momenta. It
appeared that this limit cannot be uniquely defined (it depends on the
path followed to reach the zero momenta point) and that the results of
\cite{Pisar8,Pisar9,BaierDK1,Salce1} concerning this amplitude do not
correspond to its on-shell value, but to a different way of reaching
the limit\footnote{The reason for this is easy to understand: since
  the energy variables are discrete in the imaginary time formalism,
  the only way one can consider the ``zero momenta limit'' in this
  formalism is to set first the discrete bosonic energies to zero, and
  then take the limit of zero three momenta. This way, the external
  momenta are forced to be space-like.}.  Additionally, the conclusion
according to which the pion decay amplitude vanish above the critical
point appeared to be questionable since the physical (on-shell)
amplitude has a non vanishing limit at the critical point.

To accommodate this result with the general arguments provided in
\cite{Pisar8,Pisar9,Salce1}, one can notice that both Pisarski's
symmetry argument \cite{Pisar8,Pisar9} and Salcedo's argument
\cite{Salce1} amount to the fact that one power of the quark mass (the
mass the quarks acquire through the spontaneous breakdown of chiral
symmetry, via their coupling to the average value of the $\sigma$
field) appears in the numerator when evaluating the Dirac trace
associated to anomalous amplitudes. Therefore, since the average value
$\langle \sigma\rangle$ goes to zero in the chiral phase, the
numerator vanishes above the critical point. Implicit in the argument
is the fact that the correct dimension is provided by inverse powers
of the temperature (as opposed to powers of the quark mass), as is the
case in the imaginary time formalism at the static point. In other
words, this argument is valid only if the denominator does not vanish
when the mass goes to zero. This is precisely what fails when the
amplitude is calculated on shell. One gets as expected one power of
$m=g\langle\sigma\rangle$ in the numerator, but the denominator turns
out to be $1/mT$.

The remaining power of $m$ in the denominator indicates that the
infrared or collinear behavior of the triangle diagram worsens when
$m\to 0$. In fact, as noted in \cite{Gelis6} and \cite{KumarBVH1}, the
constituent quark mass $m$ ceases to be the relevant infrared
regulator when $m$ is smaller than $gT$, and should be replaced by a
thermal mass of order $gT$ that do not vanish in the chiral limit.
Since an additional property of fermionic thermal masses is that they
respect chiral symmetry, this thermal mass cannot appear in the
Dirac's trace.  As a consequence the result $m/mT$ obtained for the
on-shell amplitude in the bare theory becomes $m/m_{\rm th}T$ after
one has resummed the quark thermal mass $m_{\rm th}\sim gT$ (if $m\ll
m_{\rm th}$).  The consequence of this regularization is that the
resummed on-shell decay amplitude vanish in the chiral limit. In other
words, Pisarski's symmetry arguments holds for the physical amplitude
only after a proper regularization (in order to get rid of all
potential infrared or collinear singularities) has been issued.

Essential in this discussion is the influence of the kinematical
conditions for the external legs on the infrared behavior of a thermal
amplitude, since it can dramatically alter one's conclusions. The
second important point is that a calculation in the imaginary time
formalism at the static point does not give a physical amplitude.
Therefore, it would be interesting to test the second half of
Pisarski's conclusions, related to the $\pi^o\sigma\to2\gamma$
amplitude, by calculating this amplitude in the real time formalism
and studying how it depends on the kinematics (up to now, this
amplitude has only be considered at the static point). Since this
amplitude is given by a four point function, it is a very complicated
task to extract this behavior in its full generality.  There is
however a toy model in which this kind of study can be done quite
simply: the 2-dimensional linear $\sigma$ model.  Indeed, in this
model the neutral pion decays into a single photon, and the analogous
of the 4-point amplitude suggested by Pisarski would be
$\pi^o\sigma\to\gamma$, which, being a 3-point function, is rather
easy to calculate.

The present paper is devoted to an analysis of the anomalous
amplitudes in the 2-dimensional $\sigma$ model at finite temperature
and chemical potential. We consider both the pion decay
$\pi^o\to\gamma$ 2-point function, and the $\pi^o\sigma\to\gamma$
3-point function. Emphasis is put on studying how these functions
depend on the kinematics in the limit of small external momenta, near
the chiral limit ($m$ small compared to $\mu$ and $T$).

The structure of the paper is as follows. Section \ref{sec:notations}
defines the model, as well as some notations and shorthands that will
be used extensively later. In section \ref{sec:2-point}, we calculate
the amplitude for the $\pi^0\to\gamma$ decay, and reduce it to a very
compact form. In section \ref{sec:3-point}, we study the amplitude of
$\pi^0\sigma\to\gamma$. Although a priori much more involved, this
amplitude can also be reduced to a simple expression.  All our results
are expressed in terms of some function $I(K)$ defined by an integral.
Basic properties and limits of this function are derived in appendix
\ref{app:I-integral}. Finally, appendix \ref{app:integrals-relations}
derives some relations between a few integrals that appear in
intermediate stages of section \ref{sec:3-point}.

\section{Conventions and notations}
\label{sec:notations}
We consider the 2-dimensional linear $\sigma$ model
\cite{Koch1,BochkK1} with two quark flavors, in which the mesons are
coupled to quark fields as indicated by the following Lagrangian:
\begin{equation}
{\cal L}\equiv i\bar{\Psi}\,\slD\Psi -2g\bar{\Psi}(\sigma t_0
+i\mbox{\boldmath$\pi$}\cdot{\imb t}\gamma^5)\Psi\; ,
\end{equation}
where $t_0=1/2$ and ${\rm Tr}(t^at^b)=\delta^{ab}/2$.
We recall that in two dimensions the Dirac algebra is defined by the
following set of relations:
\begin{eqnarray}
&&\{\gamma^\mu,\gamma^\nu\}=2g^{\mu\nu}\; ,\nonumber\\
&&\gamma^5={1\over 2}\epsilon^{\mu\nu}\gamma_\mu\gamma_\nu\; ,
\end{eqnarray}
where $\epsilon^{\mu\nu}$ is the 2-dimensional Levi-Civita tensor,
normalized by $\epsilon^{01}=+1$. For later use, let us quote first a
generic trace formula:
\begin{equation}
{\rm Tr}(\slA\slB\slC\gamma^5\gamma^\mu)=
A\cdot B{\rm Tr}(\slC\gamma^5\gamma^\mu)
-A\cdot C{\rm Tr}(\slB\gamma^5\gamma^\mu)
+B\cdot C{\rm Tr}(\slA\gamma^5\gamma^\mu)\; .
\label{eq:trace3}
\end{equation}

In order to keep the following expressions compact, it is helpful to
define the ``dual'' of a given vector by:
\begin{equation}
\widetilde{A}^\mu\equiv \epsilon^{\mu\nu}A_\nu\; ,
\end{equation}
as well as the ``wedge product'' of two vectors:
\begin{equation}
A\wedge B\equiv\epsilon^{\mu\nu}A_\mu B_\nu\; .
\end{equation}
According to these definitions, we have the obvious relations
\begin{eqnarray}
&&\widetilde{\widetilde{A}}=A\nonumber\\
&&A\wedge B=A\cdot\widetilde{B}\nonumber\\
&&(A\wedge B)^2=(A\cdot B)^2-A^2 B^2\; .
\end{eqnarray}
When $A+B+C=0$, we have also:
\begin{equation}
A\wedge B=B\wedge C=C\wedge A\; .
\end{equation}
Finally, we have
\begin{equation}
{\rm Tr}(\slA\gamma^5\gamma^\mu)=-2\widetilde{A}^\mu\; .
\label{eq:trace1}
\end{equation}

\section{$\pi^0\to \gamma$ amplitude}
\label{sec:2-point}
\subsection{Retarded amplitude}
\begin{figure}[htb]
\centerline{\includegraphics{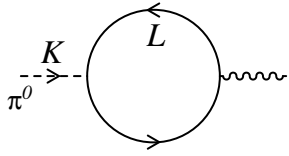}}
\caption{One-loop $\pi^0\gamma$ amplitude.}
\label{fig:2point}
\end{figure}

We consider first the 1-loop contribution to the $\pi^0\to\gamma$
decay amplitude depicted on figure \ref{fig:2point}.
The Feynman rules for the retarded-advanced formalism\footnote{One
  should pay special attention to the chemical potential. Indeed, one
  should use a chemical potential $-\mu$ in statistical weights where
  the Feynman rules give an argument $-k_0$.  In other words, $\mu$
  appears in the formalism to account for the fact that the fermions
  carry some conserved charge, and the sign of this charge for a given
  propagator depends on how one orientates the propagator.}
\cite{AurenB1,EijckW1,EijckKW1} give for the retarded amplitude the
following expression
\begin{eqnarray}
&&\Pi^\mu_{_{R}}(K)=-eg\int{{d^2L}\over{(2\pi)^2}}
{\rm Tr}\,((\slL+m)\gamma^\mu(\slL+\slK+m)\gamma^5)\nonumber\\
&&\qquad\qquad\times
\Big\{n_{_{F}}(l_0,\mu)
{\rm Disc}\Delta_{_{R}}(L)\; \Delta_{_{R}}(L+K)\nonumber\\
&&\qquad\qquad\quad +n_{_{F}}(l_0+k_0,\mu)
{\rm Disc}\Delta_{_{R}}(L+K) 
\;\Delta_{_{A}}(L)\Big\}\; ,
\label{eq:2point-expr-start}
\end{eqnarray}
where $\Delta_{_{R,A}}(L)\equiv i/(L^2-m^2\pm i l_0\epsilon)$ are the
retarded and advanced propagators, $n_{_{F}}(l_0,\mu)\equiv
1/(\exp((l_0-\mu)/T)+1)$ is the Fermi-Dirac distribution function, and
where the notation ``Disc'' denotes the discontinuity across the real
energy axis:
\begin{equation}
{\rm Disc}\Delta_{_{R}}(L)\equiv \Delta_{_{R}}(L)-\Delta_{_{A}}(L)
=2\pi\epsilon(l_0)\delta(L^2-m^2)\; .
\end{equation}

The expression of the trace is very simple
\begin{equation}
{\rm Tr}\,((\slL+m)\gamma^\mu(\slL+\slK+m)\gamma^5)=-2m\widetilde{K}^\mu \; ,
\end{equation}
and in particular it makes obvious the fact that $\Pi^\mu_{_{R}}$ is
transverse with respect to the photon momentum. Moreover, being
independent of the loop momentum $L$, it can be immediately factorized
out of the integral.

\subsection{Zero momentum limit}
At this point, it is convenient to perform the change of variable
$L+K\to -L$ on the second term of Eq.~(\ref{eq:2point-expr-start}) in
order to make the expression more symmetric. Then, the Dirac
distributions hidden in the discontinuities make the integration over
$l_0$ trivial, which gives\footnote{We have dropped the R/A
  prescription for the denominator, since it can easily be recovered
  at the very end of the calculation by substituting
  $k_0\to k_0+i0^+$.}
\begin{eqnarray}
\Pi^\mu_{_{R}}(K)=-2i m e g \widetilde{K}^\mu \int\limits_{-\infty}^{+\infty}
{{dl}\over{2\pi}} {{n_{_{F}}(-\omega_l,\mu)-n_{_{F}}(\omega_l,\mu)}
\over{2\omega_l}}
\sum_{\eta=\pm 1}{1\over{2L_\eta\cdot K+K^2}}\; ,
\end{eqnarray}
where we denote $\omega_l\equiv\sqrt{l^2+m^2}$ and $L_\eta\equiv
(\eta\omega_l,l)$. It is now trivial to perform an expansion in powers
of the external momentum $K$. The first term in this expansion, of
degree $0$ in $K$, vanish because the corresponding integrand is an
odd function of $l$. The first non vanishing term in this expansion
comes at the next order, and is of degree $1$ in $K$:
\begin{equation}
\Pi^\mu_{_{R}}(K)=i m e g \widetilde{K}^\mu I(K)\; ,
\label{eq:2point-final}
\end{equation}
where $I(K)$ is an homogeneous function of degree $0$ in $K$,
containing the non trivial part of the momentum dependence, and
defined by:
\begin{equation}
I(K)\equiv\int\limits_{-\infty}^{+\infty}{{dl}\over{2\pi}}
{{n_{_{F}}(-\omega_l,\mu)-n_{_{F}}(\omega_l,\mu)}
\over{2\omega_l}}{{K^2}\over{(L_+\cdot K)^2}}\; .
\label{eq:I-definition}
\end{equation}
This integral is studied in some important limits in the appendix
\ref{app:I-integral}.

\subsection{Discussion}
We observe for this amplitude the same features as in four dimensions.
The most striking effect is related to what happens near the chiral
limit $m\to 0$, and is visible in formulas Eqs.~(\ref{eq:I-off-shell})
and (\ref{eq:I-on-shell}). In the limit $m/T,\,m/\mu\to 0$, the
function $I(K)$ behaves as follows:
\begin{eqnarray}
&&{\rm If\ \ }K^2=0\,,\qquad 
I(K)={1\over{2\pi m^2}}\; ,\nonumber\\
&&{\rm If\ \ }K^2\not=0{\rm \ \ and \ \ }k_0\not=0\,,\qquad
I(K)={1\over{8mT}}{{k_0}\over{\sqrt{K^2}}}{1\over{\cosh^2(\mu/2T)}}\; ,
\nonumber\\
&&{\rm If\ \ }k_0=0\,,\qquad
I(K)=\left\{\matrix{& \displaystyle{{7\zeta(3)}\over{8\pi^3T^2}}
\qquad{\rm if\ \ }\mu\ll T\cr 
& \displaystyle{-{{1}\over{4\pi\mu^2}}}
\qquad{\rm if\ \ }T\ll\mu \cr}\right.\; .
\end{eqnarray}
The configuration obtained with $k_0=0$ and $\mu\ll T$ corresponds to
Salcedo's result, previously obtained in the imaginary time formalism
at the static point. This point is particular because the first term
in the expansion of $I(K)$ at small $m$ vanishes if $k_0=0$. For any
other point, the expansion starts with a term behaving like $I(K)\sim
1/mT$. However, as one turns the chemical potential on, we see that
this leading term is exponentially suppressed when $\mu\gg T$. As a
consequence, in a dense and cold system, the function $I(K)$ starts by
a term in $1/\mu^2$, whether $k_0=0$ or not.

The reason why the on-shell value is so singular when $m\to 0$ is
related to collinear singularities: in $1+1$ dimensions, all the
spatial vectors are aligned, so that we are always at the most
singular point\footnote{This is to be contrasted with what happens in
  four dimensions: there the collinear singularities are softened by
  subsequent angular integrations, so that the on-shell amplitude is
  not exceptionally singular\cite{Gelis6,GuptaN1}.}.  Here also, in
order to be able to apply Pisarski's argument, one should first
regularize the theory by resumming a thermal
mass\cite{Gelis6,KumarBVH1}. Then, all the powers of $m$ in the
denominators would be replaced by powers of the thermal mass, leaving
an uncompensated power of $m\to0$ in the numerator.  From
Eq.~(\ref{eq:2point-final}), we can write an effective Lagrangian
coupling the neutral pion to the photon:
\begin{eqnarray}
{\cal L}_{\pi^o\gamma}&&=-egm\epsilon_{\mu\nu}\int{d^2x}A^\mu(x)
I(i\partial_x)\;\partial_x^\nu\pi^o(x)\nonumber\\
&&=-eg^2\epsilon_{\mu\nu}\int{d^2x}A^\mu(x)
I(i\partial_x)\;\langle\sigma\rangle\partial_x^\nu\pi^o(x)
\end{eqnarray}
This result completes the effective coupling found by Salcedo in
\cite{Salce1}, by incorporating all the nonlocal terms. The reason why
the non-locality of this coupling has been missed in \cite{Salce1} can
be traced back in a misuse of the imaginary time techniques to get the
zero momenta limit.

\section{$\pi^0\sigma\to\gamma$ amplitude}
\label{sec:3-point}
\subsection{Retarded amplitude}
\begin{figure}[htb]
  \centerline{\includegraphics{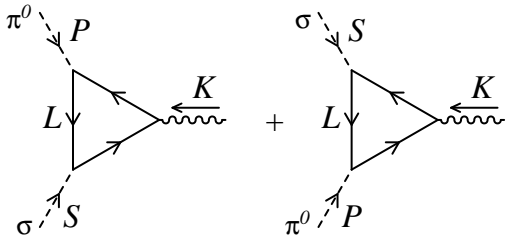}} \caption{One-loop
    $\pi^0\sigma\gamma$ amplitude.}  \label{fig:3point} \end{figure}

We now consider the one-loop contribution to the
$\pi^0\sigma\to\gamma$ amplitude, represented on figure
\ref{fig:3point}.  
In the retarded-advanced formalism, the $\Gamma_{_{ARR}}^\mu$
component of the vertex receives the following contribution from the
first diagram: \begin{eqnarray}
  &&\!\!\Gamma_{_{ARR}}^{\mu,(1)}(K,P,S)=ieg^2\int{{d^2L}\over{(2\pi)^2}}
  {\rm
    Tr}((\slL+m)\gamma^5(\slL-\slP+m)\gamma^\mu(\slL+\slS+m))\nonumber\\
  &&\qquad\qquad\times\Big\{n_{_{F}}(l_0,\mu){\rm Disc}\Delta_{_{R}}(L)
  \;\Delta_{_{R}}(L+S)\Delta_{_{A}}(L-P)\nonumber\\
  &&\qquad\qquad\quad+n_{_{F}}(l_0-p_0,\mu){\rm Disc}\Delta_{_{R}}(L-P)
  \;\Delta_{_{R}}(L)\Delta_{_{R}}(L+S)\nonumber\\
  &&\qquad\qquad\quad+n_{_{F}}(l_0+s_0,\mu){\rm Disc}\Delta_{_{R}}(L+S)
  \;\Delta_{_{A}}(L)\Delta_{_{A}}(L-P)\Big\}\; .  \end{eqnarray}
Again, the expression becomes simpler if we perform the changes of
variables $L-P\to L$ on the second term, and $L+S\to L$ on the third
one. This enables one to have common statistical weight and
discontinuity for the three terms, the tradeoff being that the trace
becomes different for the three terms. We can apply the same
manipulations to the contribution of the second diagram. With the
additional change $L\to -L$ on the second diagram, we can merge the
two contributions and find
\begin{eqnarray}
&&\Gamma_{_{ARR}}^{\mu}(K,P,S)=ieg^2\int\limits_{-\infty}^{+\infty}
{{dl}\over{2\pi}}{{n_{_{F}}(-\omega_l,\mu)-n_{_{F}}(\omega_l,\mu)}
\over{2\omega_l}}
\sum_{\eta=\pm 1}\nonumber\\ 
&&\!\!\quad\times\Big\{ {{{\rm
Tr}^\mu_a}\over{(2L_\eta\cdot K+K^2)(-2L_\eta\cdot S+S^2)}}+ {{{\rm
Tr}^\mu_b}\over{(2L_\eta\cdot P+P^2)(-2L_\eta\cdot K+K^2)}}
\nonumber\\ &&\qquad+ {{{\rm Tr}^\mu_c}\over{(2L_\eta\cdot
S+S^2)(-2L_\eta\cdot P+P^2)}} \Big\}\; , \end{eqnarray} 
where the traces are given by 
\begin{eqnarray} &&{\rm Tr}^\mu_a=-2m^2{\rm
Tr}(\slP\gamma^5\gamma^\mu) -2L_\eta\cdot S {\rm
Tr}(\slL_\eta\gamma^5\gamma^\mu) -{\rm
Tr}(\slL_\eta\slS\slK\gamma^5\gamma^\mu)\nonumber\\ &&{\rm
Tr}^\mu_b=-2m^2{\rm Tr}(\slP\gamma^5\gamma^\mu) +2L_\eta\cdot P {\rm
Tr}(\slL_\eta\gamma^5\gamma^\mu) -{\rm
Tr}(\slK\slP\slL_\eta\gamma^5\gamma^\mu)\nonumber\\ &&{\rm
Tr}^\mu_a=-2m^2{\rm Tr}(\slP\gamma^5\gamma^\mu) -{\rm
Tr}(\slS\slL_\eta\slP\gamma^5\gamma^\mu)\; .  \end{eqnarray} 

\subsection{Zero momenta limit}
We can now proceed with the expansion in powers of the external
momenta. The first term in this expansion is of degree $-1$ in the
external momenta. An explicit calculation of this term shows that it
vanishes thanks to energy-momentum conservation: $P+K+S=0$. The next
term, of degree $0$ in the external momenta, vanishes also because the
corresponding integrand is an odd function of $l$. Therefore, the
first non vanishing term is of degree $1$ in the external momenta. An
explicit extraction of this term gives
\begin{eqnarray}
&&\!\!\!\!\Gamma^\mu_{_{ARR}}(K,P,S)=i{{eg^2}\over{8}}
\int\limits_{-\infty}^{+\infty}
{{dl}\over{2\pi}}{{n_{_{F}}(-\omega_l,\mu)-n_{_{F}}(\omega_l,\mu)}
\over{2\omega_l}}
\sum_{\eta=\pm 1}\Big\{
-m^2{\rm Tr}(\slP\gamma^5\gamma^\mu)
\nonumber\\
&&\times\Big[
{{P^2}\over{(L_\eta\cdot P)^2}}{{S^2}\over{(L_\eta\cdot S)^2}}
+{{S^2}\over{(L_\eta\cdot S)^2}}{{K^2}\over{(L_\eta\cdot K)^2}}
+{{K^2}\over{(L_\eta\cdot K)^2}}{{P^2}\over{(L_\eta\cdot P)^2}}
\nonumber\\
&&
\quad+{1\over{(L_\eta\cdot K)(L_\eta\cdot P)(L_\eta\cdot S)}}
\Big(
{{K^4}\over{L_\eta\cdot K}}+
{{P^4}\over{L_\eta\cdot P}}+
{{S^4}\over{L_\eta\cdot S}}\Big)\Big]\nonumber\\
&&\quad+{{{\rm Tr}(\slL_\eta\gamma^5\gamma^\mu)}\over{L_\eta\cdot K}}
\Big[{{K^2}\over{L_\eta\cdot K}}\Big(
{{P^2}\over{L_\eta\cdot P}}-{{S^2}\over{L_\eta\cdot S}}\Big)
+{{S^4}\over{(L_\eta\cdot S)^2}}-
{{P^4}\over{(L_\eta\cdot P)^2}}\Big]\nonumber\\
&&\quad+{{{\rm Tr}(\slL_\eta\slS\slK\gamma^5\gamma^\mu)}
\over{(L_\eta\cdot K)(L_\eta\cdot S)}}
\Big[
{{S^2}\over{L_\eta\cdot S}}
-{{K^2}\over{L_\eta\cdot K}}\Big]\nonumber\\
&&\quad+{{{\rm Tr}(\slK\slP\slL_\eta\gamma^5\gamma^\mu)}
\over{(L_\eta\cdot P)(L_\eta\cdot K)}}\Big[
{{K^2}\over{L_\eta\cdot K}}
-{{P^2}\over{L_\eta\cdot P}}\Big]\nonumber\\
&&\quad+{{{\rm Tr}(\slS\slL_\eta\slP\gamma^5\gamma^\mu)}
\over{(L_\eta\cdot S)(L_\eta\cdot P)}}\Big[
{{P^2}\over{L_\eta\cdot P}}
-{{S^2}\over{L_\eta\cdot S}}\Big]\Big\}\; .
\label{eq:3point-long}
\end{eqnarray}
At this point, we can make use of Eqs.~(\ref{eq:trace3}) and
(\ref{eq:trace1}). It is now obvious that the result can be expressed
in terms of the following integrals\footnote{Note that we need only
  $J_{_{AABC}}$ when $A+B+C=0$. Having calculated $J_{_{AABC}}$ under
  this restrictive assumption, one cannot obtain $J_{_{AABB}}$ from it
  by enforcing $C=B$.}
\begin{eqnarray}
&&J_{_{AB}}\equiv\int\limits_{-\infty}^{+\infty}{{dl}\over{2\pi}}
{{n_{_{F}}(-\omega_l,\mu)-n_{_{F}}(\omega_l,\mu)}\over{2\omega_l}}
{1\over{L_+\cdot A}}{1\over{L_+\cdot B}}\nonumber\\
&&J_{_{AA}}\equiv\int\limits_{-\infty}^{+\infty}{{dl}\over{2\pi}}
{{n_{_{F}}(-\omega_l,\mu)-n_{_{F}}(\omega_l,\mu)}\over{2\omega_l}}
{1\over{(L_+\cdot A)^2}}={{I(A)}\over{A^2}}\nonumber\\
&&J_{_{AAB}}^\mu\equiv\int\limits_{-\infty}^{+\infty}{{dl}\over{2\pi}}
{{n_{_{F}}(-\omega_l,\mu)-n_{_{F}}(\omega_l,\mu)}\over{2\omega_l}}
{{L_+^\mu}\over{(L_+\cdot A)^2}}{1\over{L_+\cdot B}}\nonumber\\
&&J_{_{AABC}}\equiv m^2\int\limits_{-\infty}^{+\infty}{{dl}\over{2\pi}}
{{n_{_{F}}(-\omega_l,\mu)-n_{_{F}}(\omega_l,\mu)}\over{2\omega_l}}
{1\over{(L_+\cdot A)^2}}{1\over{L_+\cdot B}}{1\over{L_+\cdot C}}\nonumber\\
&&J_{_{AABB}}\equiv m^2\int\limits_{-\infty}^{+\infty}{{dl}\over{2\pi}}
{{n_{_{F}}(-\omega_l,\mu)-n_{_{F}}(\omega_l,\mu)}\over{2\omega_l}}
{1\over{(L_+\cdot A)^2}}{1\over{(L_+\cdot B)^2}}
\; ,
\label{eq:integrals-defs}
\end{eqnarray}
where $A,B,C$ can be any of $P,S,K$.

\subsection{Transversality}
The major difference compared to the case of the $\pi^0\gamma$
amplitude is that the above three-point amplitude is not manifestly
transverse with respect to the photon momentum. The fact that the
Dirac's traces depend upon the loop momentum $L$ indicates that it
could be necessary to calculate all the integrals before one can see
the transversality. The situation is not so intricate though, since it
happens that we only need to establish some relations between the
various integrals defined in Eqs.~(\ref{eq:integrals-defs}). In
appendix \ref{app:integrals-relations}, we show how the last three
integrals of Eq.~(\ref{eq:integrals-defs}) can be expressed as
functions of the first two. Using the relation given in
Eqs.~(\ref{eq:integral3}), (\ref{eq:integral4}) and
(\ref{eq:integral5}), as well as Eq.~(\ref{eq:JABsum}), it is a simple
matter of algebra to check the transversality of the
$\pi^0\sigma\gamma$ amplitude with respect to $K$, without the need of
calculating the various $J_{_{AA}}$ and $J_{_{AB}}$.

Once we know that the result can indeed be written as
$\Gamma^\mu=\Gamma \widetilde{K}^\mu$, one can contract the amplitude
with $S$ for instance in order to extract the coefficient $\Gamma$. A
straightforward calculation gives
\begin{equation}
\Gamma^\mu_{_{ARR}}(K,P,S)=ieg^2 \widetilde{K}^\mu F(P,S)\; ,
\end{equation}
with
\begin{equation}
F(P,S)\equiv {{P^2 S^2 I(S)+(P\cdot S) 
[P^2 I(P) + (P\cdot K) I(K)]}\over{(P\wedge S)^2}}\; ,
\label{eq:3point-final}
\end{equation}
where implicitly $K=-P-S$.  At first sight, this expression could
explode whenever two momenta become parallel. However, one can check
that this is not the case, because the numerator behaves like
$(P\wedge S)^2$ when $P\wedge S$ becomes small.

\subsection{Discussion}
We see that this 3-point amplitude involving the $\sigma$ field
depends on the same function $I(.)$ defined above, and has one power
of the mass $m$ less when compared to the $\pi^o\gamma$ amplitude, in
agreement with the general arguments of \cite{Pisar8,Pisar9,Salce1}.

Again, it is found that this limit depends on the kinematics, i.e. on
the way one is approaching the zero momenta limit. In particular, the
way this amplitude depends on $m$ at small $m$ depends strongly on the
kinematics.  This is to be contrasted with the result of
\cite{Salce1}, which only picked one particular limit.  Except at the
static point ($k_0=p_0=s_0=0$), this amplitude becomes singular at the
critical point where $m\to 0$, indicating the necessity of
regularizing the fermion propagator by a thermal mass.  After this
resummation has been performed, this amplitude is regular but does not
vanish in the chirally symmetric phase. Therefore, the conjecture of
\cite{Pisar9} holds, but only after infrared regularization.

One can also write an effective coupling associated with this amplitude:
\begin{eqnarray}
{\cal L}_{\pi^o\sigma\gamma}&&=eg^2\epsilon_{\mu\nu}\int{d^2x}
A^\mu(x)F(i\partial_{x_1},i\partial_{x_2})\nonumber\\
&&\qquad\qquad\times
[\sigma(x_2)\partial_{x_1}^\nu\pi^o(x_1)
+\pi^o(x_1)\partial_{x_2}^\nu\sigma(x_2)]
\Big|_{x_1=x_2=x}\; .
\end{eqnarray}
Of course, if one uses Eq.~(\ref{eq:limit-T0}), one finds a local
limit for this effective coupling in the limit of zero temperature and
density.

\section{Conclusions}
In this paper, we have studied the $\pi^o\gamma$ and
$\pi^o\sigma\gamma$ amplitudes in the 2-dimensional $\sigma$ model at
finite temperature and density. For both of these amplitudes, the zero
momenta limit is not unique and strongly depends on the kinematical
configuration. In particular, the imaginary time formalism should be
used with great care when looking at this limit. Indeed, if one first
sets the external discrete energies to zero, then all the information
regarding the non-locality of the amplitude is lost, and in particular
the physical limit cannot be recovered. A proper way to use the
imaginary time formalism would be to perform the sum over the loop
discrete energies while keeping nonzero external discrete energies.
After that, one should perform the analytical continuation to real
external energies, and only then consider the zero momenta limit.

The other conclusion of this work is that collinear or infrared
singularities spoil the general symmetry arguments given in
\cite{Pisar8,Pisar9,Salce1} to justify the nullity of the pion decay
into photons in the chirally symmetric phase: the {\sl on-shell} decay
amplitude in the bare theory does not vanish. For these arguments to
be valid, one should perform the resummation of a thermal mass that
will regularize the fermion propagators.

If such a regularization is used, then the conclusion is that
$\pi^o\to\gamma$ vanishes when $m\to 0$, while $\pi^o\sigma\to\gamma$
does not, in agreement with the conjecture of \cite{Pisar8}.

\section*{Acknowledgements}
We thank the Erwin Schr\"odinger Institute for support and hospitality
during the worshop ``BRST cohomology, quantization and anomalies'',
where this work started. The work of F.G. is supported by DOE under
grant DE-AC02-98CH10886.

\appendix

\section{Properties of the function $I(K)$}
\label{app:I-integral}
\subsection{Vacuum limit}
The purpose of this appendix is to study the integral $I(K)$ since all
the quantities calculated in this paper can be expressed in terms of
this function. A first check is to look at the zero temperature and
chemical potential limit of this function, which gives immediately
\begin{equation}
\lim\limits_{T,\mu\to 0^+}I(K)={1\over{2\pi m^2}}\; .
\label{eq:limit-T0}
\end{equation}
As one could expect, this $T=\mu=0$ limit does not exhibit any
non-locality in its momentum dependence, since this is a purely
thermal feature.

\subsection{Transformation into a sum}
A convenient way to look at the high temperature or density limit is
to turn the integral defining $I(K)$ in Eq.~(\ref{eq:I-definition})
into a sum, by making use of the following identity\footnote{To derive
  this formula, one can start from Mittag-Leffler's expansion of the
  $\cot$ function \cite{Chaba1,Lang1}:
\begin{equation}
\cot(z)={1\over z}+2\sum_{n=1}^{+\infty}{z\over{z^2-n^2\pi^2}}\; .
\end{equation}}:
\begin{equation}
{1\over{e^x+1}}={1\over 2}-2x\sum\limits_{n=0}^{+\infty}
{1\over{x^2+(2n+1)^2\pi^2}}\; .
\end{equation}
This identity enables one to rewrite
$n_{_{F}}(-\omega_l,\mu)-n_{_{F}}(\omega_l,\mu)$ as a series, from
which it is straightforward to first obtain
\begin{equation}
I(K)=-{1\over{\pi T^2}}{{\kappa^2+1}\over{\kappa^2-1}}
\sum\limits_{n=0}^{+\infty}\int\limits_{-\infty}^{+\infty}
dv {{v^2+\xi_+^2}\over{(v^2+\xi_-^2)^2}}
{{v^2+A_n^2}\over{(v^2+A_n^2)^2+B_n^2}}\; ,
\end{equation}
where we denote
\begin{eqnarray}
&&v\equiv {l\over T}\; ,\qquad
\kappa\equiv k_0/k\; ,\qquad 
\xi_\pm^2\equiv{{m^2}\over{T^2}}{{\kappa^2}\over{\kappa^2\pm 1}}\nonumber\\
&&A_n^2\equiv\pi^2(2n+1)^2+{{m^2}\over{T^2}}-{{\mu^2}\over{T^2}}\; ,\qquad
B_n^2\equiv4\pi^2(2n+1)^2 {{\mu^2}\over{T^2}}\; .
\end{eqnarray}
At this stage, it remains to perform term by term the integration over
$dv$, which is elementary and yields
\begin{eqnarray}
&&\int\limits_{-\infty}^{+\infty}\!\!
dv {{v^2+\xi_+^2}\over{(v^2+\xi_-^2)^2}}
{{v^2+A_n^2}\over{(v^2+A_n^2)^2+B_n^2}}
=-{\pi\over{2\xi_-}}{{\partial}\over{\partial\xi_-}}\left[
\vphantom{{{\sqrt{A_n^4+B_n^2}+A_n^2}\over{2(A_n^4+B_n^2)}}}
{1\over{\xi_-}}
{{(\xi_+^2-\xi_-^2)(A_n^2-\xi_-^2)}\over{(A_n^2-\xi_-^2)^2+B_n^2}}
\right.\nonumber\\
&&\qquad\qquad\qquad\qquad
+{{A_n^4+B_n^2-2\xi_+^2A_n^2+(\xi_+^2-\xi_-^2)\sqrt{A_n^4+B_n^2}
+\xi_+^2\xi_-^2}\over{(A_n^2-\xi_-^2)^2+B_n^2}}
\nonumber\\
&&\qquad\qquad\qquad\qquad\qquad\times\left.
\sqrt{{\sqrt{A_n^4+B_n^2}+A_n^2}\over{2(A_n^4+B_n^2)}}\;
\right]\; .
\end{eqnarray}

\subsection{Chiral limit far from the light cone}
Another interesting limit is the chiral limit, where $m/T$ goes to
zero, while $\mu/T$ is kept fixed. It is very easy to extract from the
above formula a systematic expansion in powers of $m/T$. It is just a
matter of expanding at small $\xi_\pm$ the above expression, which
gives for the first two orders:
\begin{eqnarray}
I(K)={1\over{mT}}\left[{{k_0}\over{\sqrt{K^2}}}F_0\left({\mu\over T}\right)
+{m\over T}{{K^2+2k^2}\over{K^2}}F_1\left({\mu\over T}\right)
+{\cal O}\left({m^2\over T^2}\right)
\right]\; ,
\label{eq:I-off-shell}
\end{eqnarray}
where the coefficients are given by
\begin{eqnarray}
&&F_0\left({\mu\over T}\right)={1\over{\pi^2}}{\rm Re}
\sum\limits_{n=0}^{+\infty}
{1\over{(2n+1+i\mu/\pi T)^2}}\; ,
\nonumber\\
&&F_1\left({\mu\over T}\right)={1\over{\pi^3}}\sum\limits_{n=0}^{+\infty}
(2n+1){{(2n+1)^2-3\mu^2/\pi^2 T^2}\over{((2n+1)^2+\mu^2/\pi^2 T^2)^3}}
\; .
\end{eqnarray}
Introducing the ``digamma'' function
\begin{equation}
\psi(z)\equiv {d\over{dz}}\ln\Gamma(z)
\end{equation}
and a series representation of its first derivative
\begin{equation}
\psi^\prime\left({{1+z}\over 2}\right)=4\sum\limits_{n=0}^{+\infty}
{1\over{(2n+1+z)^2}}\; ,
\end{equation}
it is immediate to check\footnote{The equality on the first line is
  exact and comes from the formula \cite{Chaba1,Lang1}
\begin{equation}
\Gamma(z)\Gamma(1-z)={\pi\over{\sin(\pi z)}}\; ,
\end{equation}
while the limit $\mu \gg T$ in the last line is obtained from
Stirling's expansion \cite{Chaba1,Lang1} for $\Gamma(z)$.  }
\begin{eqnarray}
&&F_0\left({\mu\over T}\right)={1\over{8\pi^2}}\left[
\psi^\prime\left({1\over 2}+i{{\mu}\over{2\pi T}}\right)
+
\psi^\prime\left({1\over 2}-i{{\mu}\over{2\pi T}}\right)
\right]={1\over{8\cosh^2(\mu/2T)}}\; ,\nonumber\\
&&F_1\left({\mu\over T}\right)=
-{1\over{32\pi^3}}\left[
\psi^{\prime\prime}\left({1\over 2}+i{{\mu}\over{2\pi T}}\right)
+
\psi^{\prime\prime}\left({1\over 2}-i{{\mu}\over{2\pi T}}\right)
\right]\nonumber\\
&&\qquad\quad\;\;
\approx\left\{\matrix{& \displaystyle{{7\zeta(3)}\over{8\pi^3}}
\qquad{\rm if\ \ }\mu\ll T\cr 
& \displaystyle{-{{T^2}\over{4\pi\mu^2}}}
\qquad{\rm if\ \ }T\ll\mu \cr}\right.\; .
\end{eqnarray}

\subsection{Light cone limit}
Note however that this expansion is not valid near the light cone.
Indeed, its derivation assumed that a small $m/T$ would imply a small
$\xi_-$, which is not true if $K^2$ is small (or equivalently
$\kappa^2\approx 1$).  Sufficiently close to the light cone, $\xi_-$
becomes large and a different kind of expansion must be considered. In
this region of phase space, one can write:
\begin{eqnarray}
\lim\limits_{K^2\to 0}I(K)&&=\lim\limits_{\xi_-\to +\infty}
{1\over{2\xi_- T^2}}
{2\over{\kappa^2-1}}
\sum\limits_{n=0}^{+\infty}{1\over{[\xi_-+(2n+1)\pi]^2}}
\nonumber\\
&&=\lim\limits_{\xi_-\to +\infty}
{{\xi_-}\over{m^2}}
{1\over {4\pi^2}}\psi^\prime\left({1\over 2}+{{\xi_-}\over{2\pi}}\right)
\; .
\end{eqnarray}
where we have used $(\kappa^2-1)^{-1}\approx\xi_-^2 T^2/m^2$.  Making
use of Stirling's formula \cite{Chaba1,Lang1}, we obtain:
\begin{equation}
\lim\limits_{K^2\to 0} I(K)={1\over{2\pi m^2}}\; .
\label{eq:I-on-shell}
\end{equation}
We notice that the on-shell value of $I(K)$ is totally immune to
corrections due to temperature or density. The origin of this property
can be understood from Eq.~(\ref{eq:I-definition}): when $K^2=0$, the
denominator behaves like $k^2 m^4/l^4$ (this is reminiscent of a
collinear singularity cured by the mass $m$), and the integral is
completely dominated by its ultraviolet sector. As a consequence,
$I(K)$ is saturated by the vacuum contribution for this value of $K^2$.
Away from the light cone, thermal corrections of order $1/mT$ appear
in $I(K)$.

\section{Relations between some integrals}
\label{app:integrals-relations}
In this appendix, we establish some useful relations between the five
integrals defined in Eq.~(\ref{eq:integrals-defs}). Let us start by
the study of $J^\mu_{_{AAB}}$: this integral satisfies the $2\times 2$
linear system
\begin{equation}
\left\{\matrix{
A_\mu J^\mu_{_{AAB}}&=J_{_{AB}}\vphantom{\Big)}\cr
B_\mu J^\mu_{_{AAB}}&=J_{_{AA}}\vphantom{\Big)}\; ,\cr}\right.
\end{equation}
the resolution of which gives the two components of $J^\mu_{_{AAB}}$
as functions of $J_{_{AA}}$ and $J_{_{AB}}$:
\begin{equation}
J^\mu_{_{AAB}}={{\widetilde{B}^\mu J_{_{AB}}-\widetilde{A}^\mu J_{_{AA}}}
\over{A\wedge B}}\; .
\label{eq:integral3}
\end{equation}
In order to obtain $J_{_{AABB}}$ it is convenient to define first the
second rank tensor
\begin{equation}
J^{\mu\nu}_{_{AABB}}\equiv \int\limits_{-\infty}^{+\infty}{{dl}\over{2\pi}}
{{n_{_{F}}(-\omega_l,\mu)-n_{_{F}}(\omega_l,\mu)}\over{2\omega_l}}
{{L_+^\mu}\over{(L_+\cdot A)^2}}{{L_+^\nu}\over{(L_+\cdot B)^2}}\; ,
\end{equation}
from which we can obtain $J_{_{AABB}}$ as
$g_{\mu\nu}J^{\mu\nu}_{_{AABB}}$. The three independent components of
this symmetric tensor can be obtained via the resolution of the
following $3\times 3$ linear system
\begin{equation}
\left\{\matrix{
A_\mu B_\nu J^{\mu\nu}_{_{AABB}}&=J_{_{AB}}\vphantom{\Big)}\cr
A_\mu A_\nu J^{\mu\nu}_{_{AABB}}&=J_{_{BB}}\vphantom{\Big)}\cr
B_\mu B_\nu J^{\mu\nu}_{_{AABB}}&=J_{_{AA}}\vphantom{\Big)}\; ,\cr
}\right.
\end{equation}
which finally gives
\begin{equation}
J_{_{AABB}}={{2(A\cdot B)J_{_{AB}}-A^2 J_{_{AA}} -B^2 J_{_{BB}}}
\over{(A\wedge B)^2}}\; .
\label{eq:integral4}
\end{equation}
The same method can be applied to $J_{_{AABC}}$, which gives:
\begin{equation}
J_{_{AABC}}={{A^2J_{_{AA}}+B^2 J_{_{AB}}+C^2 J_{_{AC}}}
\over{(A\wedge B)^2}}\; ,
\label{eq:integral5}
\end{equation}
under the assumption that $A+B+C=0$. We also need the following relation
\begin{equation}
J_{_{AB}}+J_{_{BC}}+J_{_{CA}}=0\; ,
\label{eq:JABsum}
\end{equation}
which is valid when $A+B+C=0$.


\end{document}